\documentclass[a4paper,12pt]{article}
\usepackage[hidelinks]{hyperref}
\usepackage{graphicx}
\usepackage{setspace} 
\usepackage{geometry}
\usepackage{array}
\usepackage{multirow}
\usepackage{bm}

\begin{document}
\thispagestyle{empty}
\begin{center}
  \large{
    {\bf Faculty Induction Programme (FIP-3)\\~\\~\\ 
      1st June 2022 -- 2nd July 2022}\\~\\~\\~\\
    {\em Topic: Effect of Spreading Knowledge Centers --
      A Physics-based Approach}\\~\\~\\~\\
    Submitted as partial fulfilment of UGC sponsored\\ 
    Faculty Induction Programme\\~\\~\\~\\
    {\em By\\
      Saurish Chakrabarty\\
      Department of Physics\\
      Acharya Prafulla Chandra College\\ 
      West Bengal State University}
    \vfill
    \includegraphics[width=1in]{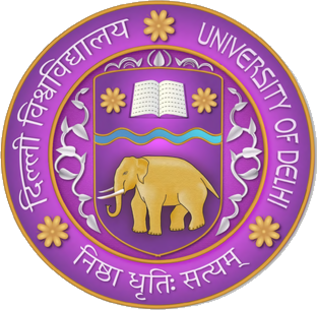}\\~\\
    {\bf CENTRE FOR PROFESSIONAL DEVELOPMENT IN HIGHER EDUCATION\\ 
      UNIVERSITY OF DELHI\\
      DELHI-110007}}
\end{center}
\clearpage\pagebreak

\begin{center}
  Declaration of Originality 
\end{center}

\noindent
Name: Saurish Chakrabarty\\
Subject: Physics\\
College: Acharya Prafulla Chandra College\\
University: West Bengal State University\\~\\

\noindent
Title of the paper: Effect of Spreading Knowledge Centers -- A Physics-based Approach\\~\\

\noindent
I hereby declare that the project report/paper titled
``Effect of Spreading Knowledge Centers -- A Physics-based Approach''
that I am submitting to fulfill the requirement towards completion
of my Faculty Induction Programme (FIP-3)
(1st June 2022  2nd July 2022), at CPDHE, University of Delhi,
represents my original work and that I have used no other
sources except those that I have indicated in the references.
All data, tables, figures and text citations which have been
reproduced from any other source, including the internet,
have been explicitly acknowledged as such. I am aware of the
consequences of non-compliance of my above declaration.\\~\\

\noindent
Saurish Chakrabarty\\~\\
Kolkata, West Bengal\\~\\
Date: June 20, 2022

\clearpage\pagebreak
\doublespacing
\begin{center}
  \Large{\textsc{Abstract}}
\end{center}
We use a simple physics-inspired model to get an idea about how to
enhance the speed with which a society becomes
educated if we strategically place our knowledge spreading centers
(teachers or educational institutions).
We study knowledge spreading using the Ising model, a well-studied
model used in physics, specifically statistical mechanics,
to describe the phenomenon of ferromagnetism.
In the social context, up and down spins are mapped to knowledgeable and ignorant
individuals.
We introduce some knowledgeable individuals into an otherwise ignorant society and see
how their number increases with time, when evolved using the Metropolis algorithm.
We find that the knowledge of the society grows faster when the initial group of
knowledgeable individuals is maximally spread out.
We quantify this effect using the doubling time and look at the distribution of the
doubling time as a function of ``temperature''.
In the social context, the energy is identified as the (lack of) happiness of neighbours
and temperature is a parameter that quantifies how important happiness is
in the society.
We point out several limitations
of this study in order to facilitate future research.

\vspace{1in}

\begin{center}
  \textsc{Keywords}\\
  \noindent
  Ising model, knowledge spreading, doubling time,
  sociophysics, econophysics
\end{center}

\clearpage\pagebreak
\section{Background and Introduction}
The Ising model \cite{ising25} was introduced as a simple model
by Lenz in 1920 to explain ferromagnetism, a phenomenon in which
below a certain critical temperature, local microscopic
magnetic moments spontaneously align to give rise to a macroscopic
magnetization. 
In this work, we use this model in a social setting where the
Ising model can be seen as a model that describes a society which
veers towards an equilibrium state where knowledge that is initially
available to a small set of individuals in a population,
later gets distributed among a large fraction of the population.
We begin with a brief review of the Ising model in
the standard statistical mechanistic setting.
This model is described on a $d$-dimensional regular lattice with ``spins'',
$\left\{S_i\right\}$,
residing on every lattice site.
This work focuses on $d=2$ (two spatial dimensions).
Each spin can take one of two values, $\pm1$, (``up'' and ``down'').

Applications of the Ising model and its variants to social settings are
not new.
For a short but excellent review see Ref. \cite{stauffer08} and the
references therein.
For another, more exhaustive review of physics-inspired models
in financial economics, including the Ising model, see Ref. \cite{sornette14}.
The Ising model has been used in the past to understand damage spreading.\cite{svenson02}
It has also been used to understand patterns in tax evasion.\cite{zaklan09}
In this context, the up and down spins were used to
represent ``honest'' tax payers and ``cheaters''.
Self organization in the financial markets can also be studied using the
Ising model.\cite{zhou07}
In this work, stochastic evolution was
used instead of the standard Hamiltonian formulation.
Talking about self organization, it is important to point out that
self organization, especially the phenomenon self organized criticality,
is a very important and interesting branch of the statistical mechanics
of complex systems and explains a lot of universality that we see in nature.\cite{bak88}
A model based on the Ising model has been used to understand the
role of social impact.\cite{kohring96}
Here, the Ising variables represented the ``opinions'' and
there were two kinds of coupling constants ``persuasiveness'' and ``support''.

The Ising model is described by the following simple Hamiltonian.
\begin{eqnarray}
  {\cal H}=-J\sum_{\langle i,j\rangle}S_iS_j-h\sum_iS_i
  \label{isingHam}
\end{eqnarray}
[For a reader unaware of what the Hamiltonian is,
it represents the energy function. When thermal effects are not important, a system minimizes
its energy, and when they are important, there is a balance between minimization of the
energy and maximization of the entropy (tendency to approach a distribution in which the number
of microscopic configurations is high, $i.e.$, the system is confused).]
The coupling constant $J$ is taken to be positive in the ferromagnetic Ising model and
sets the scale of energies.
The first term is a sum over nearest neighbours
(denoted by the angular brackets) and to satisfy this term,
nearest neighbours must have the same spin.
Here, the word satisfy refers to allowing it to contribute
the lowest possible energy.
When $h=0$, the model results in a ground state where all the
spins point in the same direction.
When $h\neq0$, in the ground state, all the spins take the sign of $h$.
The constant, $h$, therefore, plays the role of a magnetic field and spins align in the direction
of the magnetic field.
Generalizations of the two constants give rise to many interesting phenomena
such as antiferromagnetism, spin glasses and localization-delocalization transitions.

Equilibrium properties of the Ising model at some external temperature, $T$,
(canonical ensemble)
are obtained by calculating the partition function,
\begin{eqnarray}
  Z=\mbox{Tr}~e^{-\beta{\cal H}},
\end{eqnarray}
where $\beta=\frac{1}{k_BT}$, $k_B$ being the Boltzmann constant
and the symbol $\mbox{Tr}$ (trace) represents the sum over all possible
configurations.

The partition function for the Ising model was calculated by Ising in
one spatial dimension \cite{ising25} and by Onsager in two spatial
dimensions \cite{onsager44}.
Onsager used the well-known transfer matrix method
that was invented a few years ago and played a role in popularizing the method
which later proved to be a useful tool in solving many problems in
all areas of physics, including statistical mechanics,
particularly in the area of complex systems.
It is generally believed that a closed-form analytical solution
of the three-dimensional Ising model is not possible.
In two and more dimensions, the Ising model shows a transition
from a low-temperature ferromagnetic phase (most spins pointing in the same direction)
to a high-temperature paramagnetic phase (no preferred direction of the spins).
This transition occurs at a critical temperature, $T_c$, that depends on the
number of spatial dimensions.

\begin{figure}
  \centering
  \vspace{-1cm}
  \includegraphics[width=\textwidth]{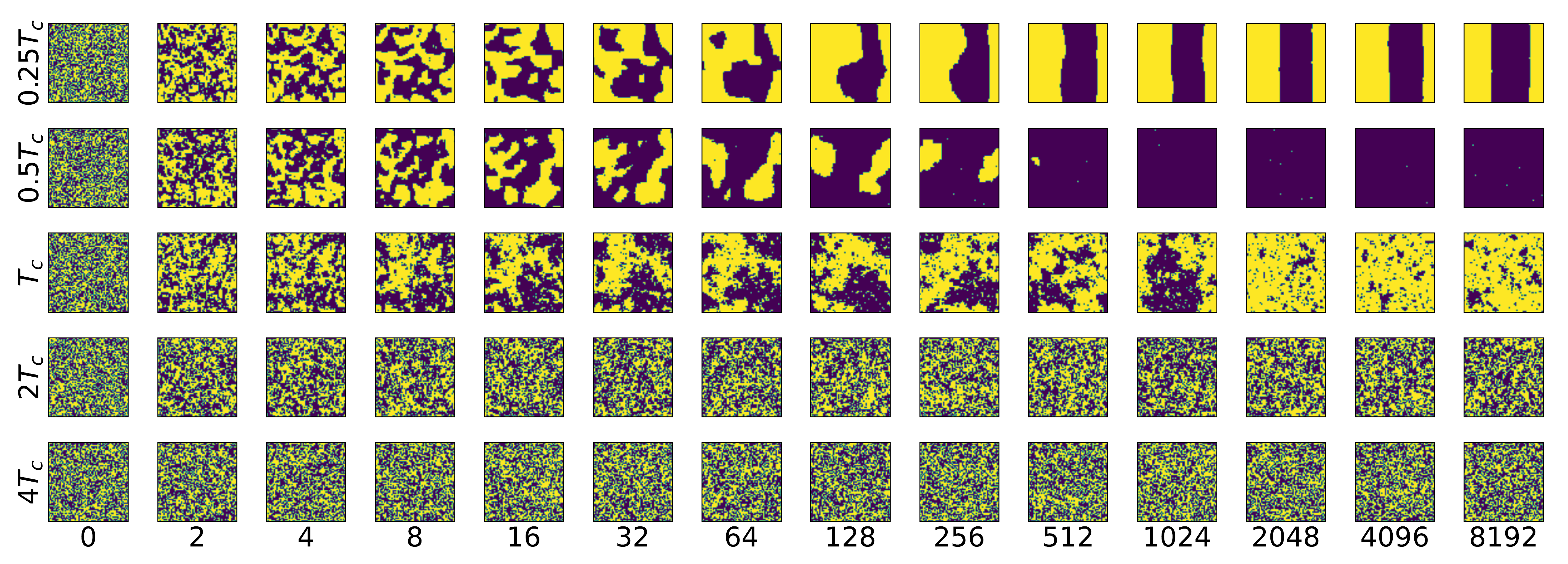}
  \caption{Configurations of the two-dimensional Ising model with periodic boundary conditions,
    evolved using the Metropolis algorithm,
    starting from a random configuration,
    with $J=1$ and $h=0$.
    Each row corresponds to a fixed temperature that is written on the left of the row.
    The number of Monte Carlo cycles increases from left to right and is written at the bottom of
    each column.
    For the two-dimensional Ising model, $T_c\approx2.269J/k_B$ and the
    equilibrium configurations below this temperature are ferromagnetic and
    above this temperature are paramagnetic.
    As is clear from the data, in the ferromagnetic phase,
    presence of domain walls across the system costs very little
    energy and can last very long.
  }
  \label{isingConfRand}
\end{figure}

\begin{figure}
  \centering
  \includegraphics[width=\textwidth]{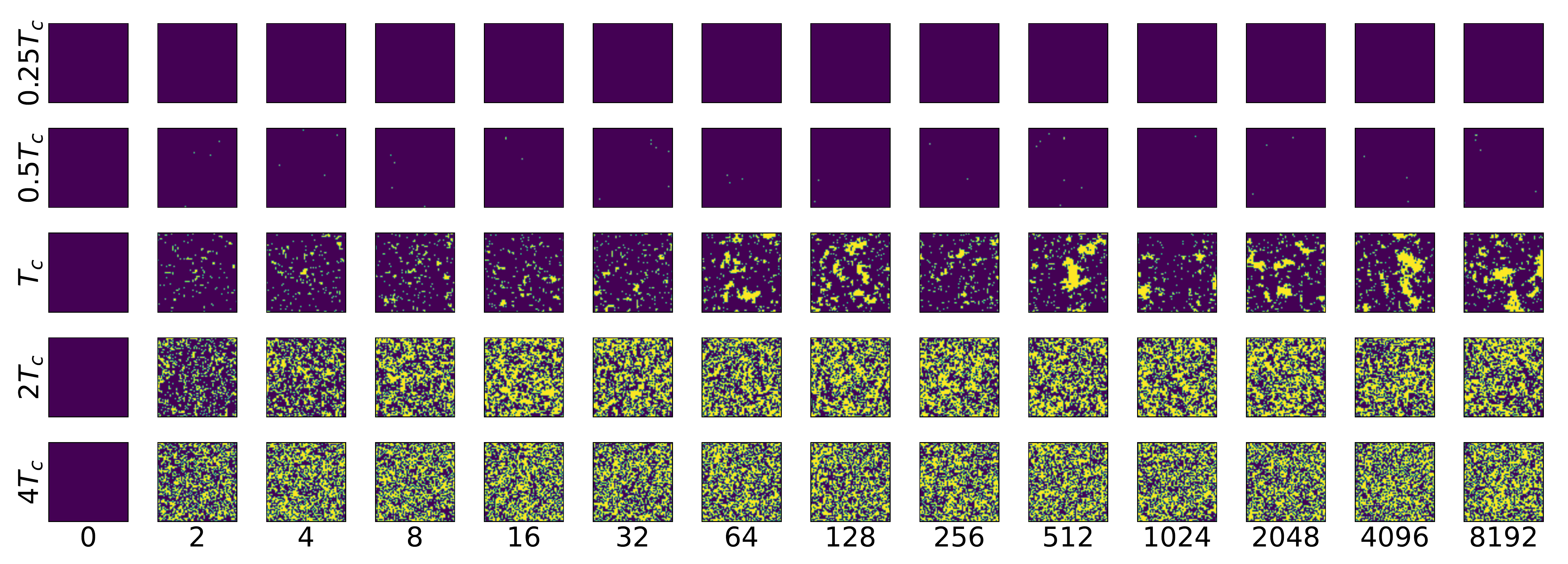}
  \caption{Configurations starting from a configuration with all down spins.
    All other details are the same as Fig. \ref{isingConfRand}.
    Ignoring the vertical domain wall at the lowest temperature in
    Fig. \ref{isingConfRand} and with the up-down symmetry in the Hamiltonian when $h=0$,
    the configurations on the last column are statistically similar to the corresponding
    figures on the last column of Fig. \ref{isingConfRand}.
  }
  \label{isingConfZero}
\end{figure}

Configurations corresponding to the equilibrium distribution at a given
temperature can be sampled using the Metropolis algorithm.\cite{metropolis}
In this method, a spin is chosen at random and flipped. If this causes
a reduction in the energy, then the move is accepted. If, on the other
hand there is a rise in the energy, then the move is accepted with
probability, $e^{-\beta\Delta E}$, where $\Delta E$ is the increase in energy.
This process results in a distribution of configurations such that the
probability of having a configuration with energy $\cal H$ is
$\frac{1}{Z}e^{-\beta{\cal H}}$.
Configurations obtained by evolution using this algorithm are shown
in Figs. \ref{isingConfRand} and \ref{isingConfZero}.
In Fig. \ref{isingConfRand}, we have chosen a random initial configuration, $i.e$,
each spin can be either up or down, with equal probability.
In Fig. \ref{isingConfZero}, we start with initial configurations
in which all spins are down.

\section{Application to Knowledge Spreading}
The Ising model can be used to describe any two-level system.
In the application we have in mind, it will be convenient to work with
variables which take values in $\{0,1\}$ instead of $\{-1,+1\}$.
This change is achieved by the following change of variables.
\begin{eqnarray}
  \sigma_i=\frac{S_i+1}{2}\mbox{, or equivalently, }
  S_i=2\sigma_i-1
\end{eqnarray}

We apply the Ising model to the following social scenario.
Consider a society with $N=L\times L$ individuals,
each occupying a site
on a square grid of side $L$.
(These $N$ sites could be occupied by $N$ households as well.)
Each person is either aware or unaware of some fact, $i.e$.,
can be in two states -- one or zero (up or down).
The person is ``happy'' if surrounded by people in the same state.
We can characterize each connection (bond) with its happiness.
Lower the energy, higher is the happiness.
The individuals (or bonds) keep evolving (fighting or struggling)
in the quest for happiness.

It is clear that the above setup has the ingredients to be modeled using the
Ising model with zero field. Each person has a knowledge level $\sigma\in\{0,1\}$
and the Hamiltonian $\cal H$ in Eq. \ref{isingHam}, can be thought to
represent the negative of happiness ($i.e.$, sadness).
The environment introduces fluctuations and
enforces some of their statistical properties.
These are controlled by the ``temperature'', $T$.
At a low temperature, the society veers towards a state with lowest energy
(maximum happiness).
At extremely high temperatures, the system tends to become
maximally disordered (maximum entropy).
In this situation, states of neighbours are uncorrelated (people do not care
about each other and are not affected by their state). Happiness in the society is not important.

Another interesting point to note is that the situation in which every person in the society is
knowledgeable has the same total happiness as the situation in which no person is.
More generally, there is a symmetry that flipping the states of all the individuals
does not affect the society's overall energy or happiness.

The above description is incomplete without the specification of the boundary conditions.
The society has periodic boundary conditions. The exit at the north boundary
is the entry through the south boundary and vice versa.
The same is true for the east and west boundaries.
This kind of boundary condition has numerical advantages and finite size effects are small.
In addition, all sites are equivalent.
Everyone has the same number of neighbours.
No location can be called a central location and
no location can be called a boundary.
In what follows, we will destroy this beautiful spatial homogeneity.

\begin{figure}
  \centering
  \includegraphics[height=0.2\textwidth]{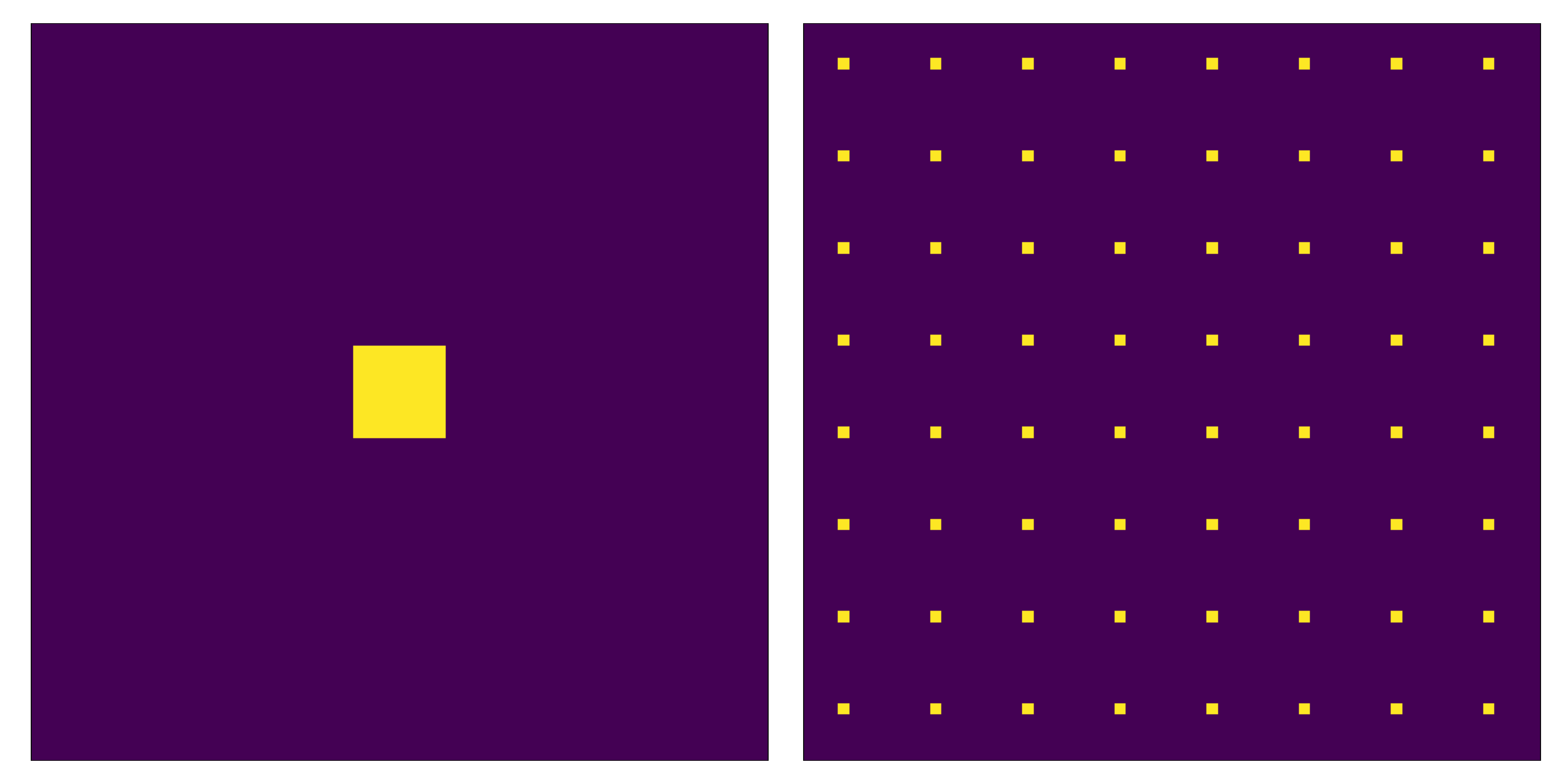}
  \caption{The two chosen starting configurations --
    {\em centered} (left) and {\em distributed}
    (right), shown here with $n_s=64$.}
  \label{starts}
\end{figure}

We now consider the scenario in which $n_s=l_s\times l_s$ knowledgeable people are
introduced into the society in which everyone is ignorant.
These $n_s$ individuals stay knowledgeable forever.
We look at the effect of their locations on the future knowledge
content of the society.
We evolve the society using the Metropolis algorithm and calculate the time (Monte
Carlo steps) taken to double the number of knowledgeable individuals.
We denote this by time by $\tau$.
One Monte Carlo step corresponds to $N$ spin-flip (knowledge-flip) attempts.
We compare the following two situations.
\begin{enumerate}
\item In the beginning all the knowledgeable individuals stay close by. They occupy all sites in a
  $l_s\times l_s$ square somewhere in the system, as shown in the left panel of Fig. \ref{starts}.
\item In the beginning all the knowledgeable individuals stay in a regular array,
  as far from each other as possible, as shown in the right panel of Fig. \ref{starts}.
\end{enumerate}

\begin{figure}
  \centering
  \includegraphics[height=0.5\textwidth]{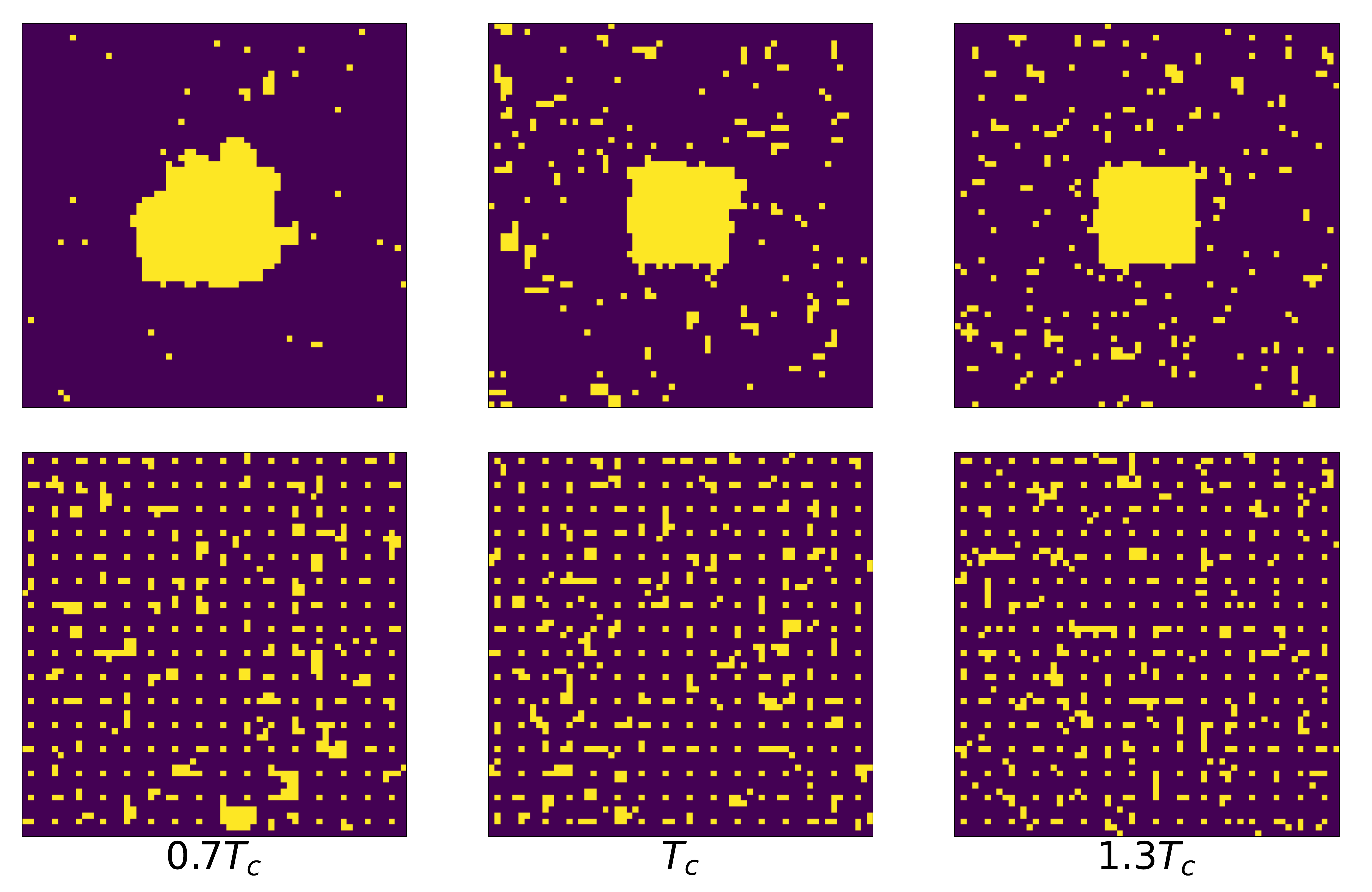}
  \caption{
    Configurations after the doubling time, $\tau$.
    The two top and the bottom rows correspond to
    {\em centered} and {\em distributed}
    starting configurations, respectively.
    The configurations shown are for $n_s=256$.
    Each column correspond to a fixed temperature that is
    mentioned at the bottom.
  }
  \label{sends}
\end{figure}

\section{Details of the Study and Results}
We focus on a system with $N=4096$ sites.
The initial number of knowledgeable individuals, $n_s$, was taken from the set $\{16,64,256\}$.
The temperature, $T$, was chosen from the set $\{0.7T_c, T_c, 1.3T_c\}$.
The choice of these parameters were determined only by the time and the computational
resources that were available for doing the work. More of these parameter values
should be explored in future.

In Fig. \ref{sends}, we see the temperature dependence of the
pattern in which knowledge spreads in the above
two situations.
Larger clusters of knowledge centers are seen at lower temperatures.
The starting configuration had $n_s=256$ for the snapshots shown in this figure.

\begin{figure}
  \centering
  \includegraphics[height=0.25\textwidth]{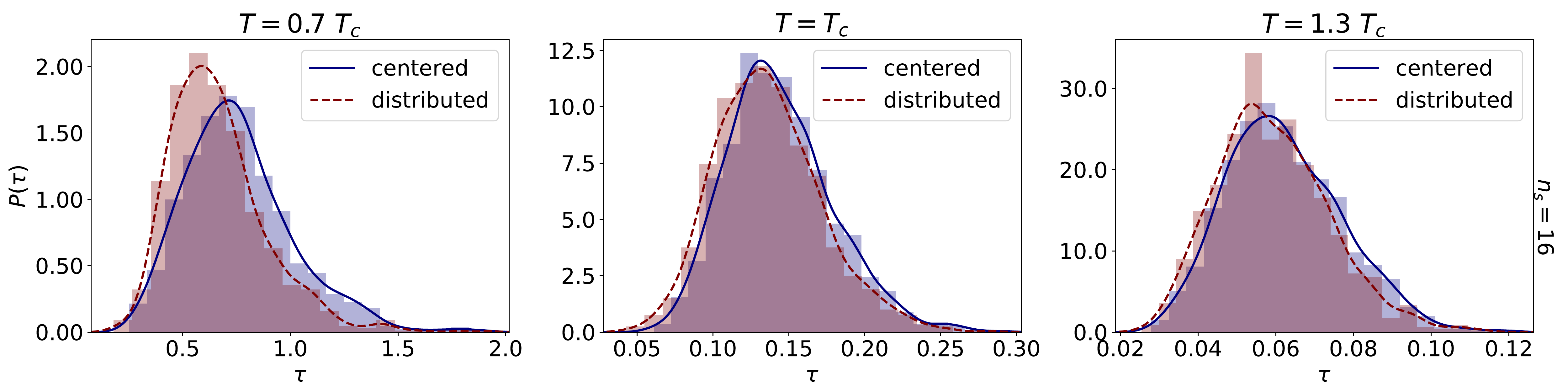}\\
  \includegraphics[height=0.25\textwidth]{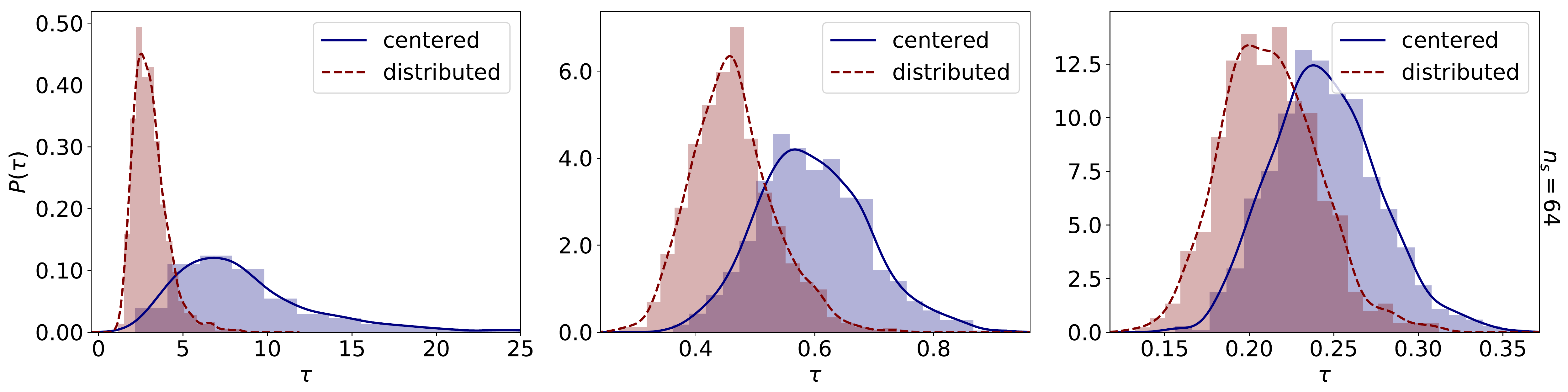}\\
  \includegraphics[height=0.25\textwidth]{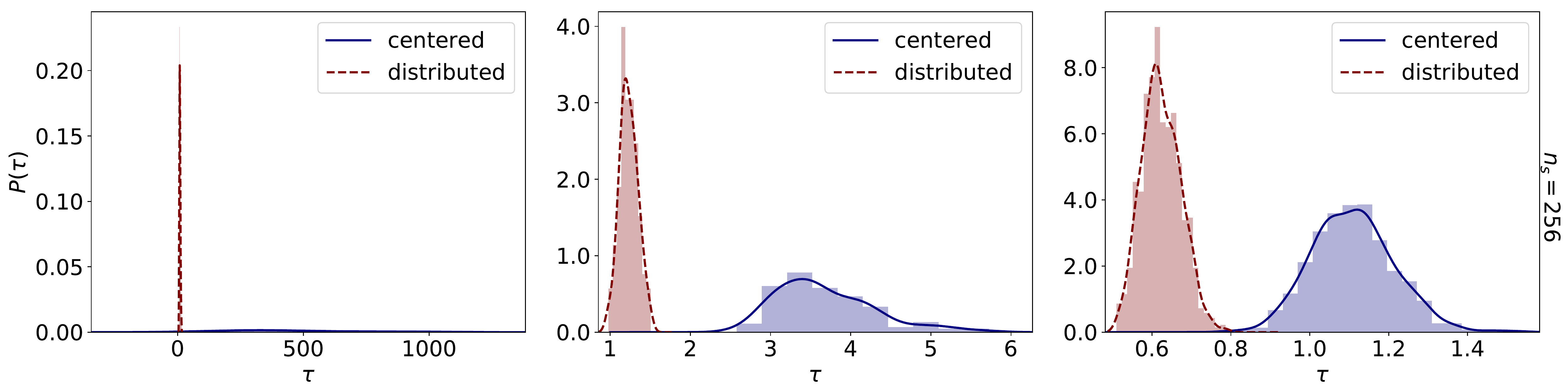}
  \caption{
    The distributions of the doubling time, $\tau$.
    In the top row, there are $n_s=16$ knowledgeable people initially.
    In the middle row, $n_s=64$. 
    Finally, in the bottom row, $n_s=256$.
    The three columns are for three different temperatures (mentioned at
    the top of each column).
    In each of the three panels, centered and distributed starting seeds have
    been shown in different colors. The lines are obtained via Kernel
    Density Estimation (computed using the $pandas$ module in $python$
    \cite{pandasKDE}).
    The choice of this method was done arbitrarily with a purpose to
    have a guiding trend line for the reader (any other spline algorithm
    would have also served our purpose and due to the limited size of the
    data set used,
    features other than the mean
    and variance of the data should not be taken too seriously). 
  }
  \label{t2}
\end{figure}

In Fig. \ref{t2}, we see that when $n_s=16$, the locations and spread of the distributions
are affected slightly by the choice of the starting configuration, the blue-solid and 
red-dashed lines representing the initially centered and spread-out knowledge centers
respectively.
The distribution with the spread-out initial configuration is narrower and
peaked at a lower time.
The effect becomes less significant with increasing temperature.
If we look at higher values of $n_s$, the effects become more pronounced.
These details are further supported by the numbers shown in Table \ref{t2data}.
The means and standard deviations of the data sets are mentioned here.
More importantly, the sizes of the data sets are also given.
\begin{table}
  \centering
  \begin{tabular}{|c|c|c|c|c|c|}
    \hline
    $\bm{n_s}$ & $\bm{T}$ & \begin{minipage}{1.5in}{\tiny{~\\}}\bf Initial location of knowledge centers\end{minipage} & $\bm{\langle\tau\rangle}$ & $\bm{\sigma_\tau}$ & \begin{minipage}{1in}{\bf Size of data set}\end{minipage}
    \\
    \hline
    \multirow{6}*{16} & \multirow{2}*{0.7$T_c$} & centered & 0.758 & 0.252 & 1000\\
                      &  & distributed & 0.662 & 0.222 & 1000\\
    \cline{2-6} & \multirow{2}*{$T_c$} & centered & 0.144 & 0.0348 & 1000\\
                      &  & distributed & 0.136 & 0.0347 & 1000\\
    \cline{2-6} & \multirow{2}*{1.3$T_c$} & centered & 0.0622 & 0.0149 & 1000\\
                      &  & distributed & 0.0593 & 0.0144 & 1000\\
    \hline
    \multirow{6}*{64} & \multirow{2}*{0.7$T_c$} & centered & 8.66 & 4.35 & 1000\\
                      &  & distributed & 3.07 & 1.02 & 1000\\
    \cline{2-6} & \multirow{2}*{$T_c$} & centered & 0.597 & 0.0919 & 1000\\
                      &  & distributed & 0.462 & 0.0687 & 1000\\
    \cline{2-6} & \multirow{2}*{1.3$T_c$} & centered & 0.245 & 0.0316 & 1000\\
                      &  & distributed & 0.212 & 0.0284 & 1000\\
    \hline
    \multirow{6}*{256} & \multirow{2}*{0.7$T_c$} & centered & 442 & 260 & 38\\
                      &  & distributed & 8.67 & 1.97 & 100\\
    \cline{2-6} & \multirow{2}*{$T_c$} & centered & 3.67 & 0.604 & 142\\
                      &  & distributed & 1.23 & 0.112 & 100\\
    \cline{2-6} & \multirow{2}*{1.3$T_c$} & centered & 1.11 & 0.105 & 1000\\
                      &  & distributed & 0.624 & 0.0489 & 1000\\
    \hline
  \end{tabular}
  \caption{Information about the data set generated from simulation: Means and standard deviations of
    the doubling times are shown in addition to the sizes of the data set used (which are small for
    some sets of parameters).}
  \label{t2data}
\end{table}

\section{Conclusion}
To conclude,
we see that a simple physics-inspired model can be used to
give us an idea about how to enhance the speed with which a society becomes
educated. A good strategy adopted to place our knowledge spreading centers
(teachers) plays an important role.
In an alternate interpretation in which the sites in our model are occupied by households,
the initial centers of knowledge may represent educational institutions,
libraries, news agencies or any other institution depending on the kind of
knowledge we wish to focus on.
Our results suggest that a good strategy is to spread out the knowledge centers
as much as possible.
Environmental factors (here ``temperature'') play a big role.
When the temperature is low, the choice of strategy has a much profound effect, $i.e.$,
it is significantly better to spread out the teachers.
The difference gets washed out with ``thermal'' fluctuations.
Lastly, if we have more resources, it is even more fruitful, and therefore important, to
adopt the better strategy of spreading out our resources.

\section{Limitations and Possible Extensions}
This work was done as a part of a Faculty Induction Programme and very
little time could be spent on it.
As such, the following areas
should be explored more exhaustively in order to establish the 
universal applicability of the results.
\begin{enumerate}
\item System size dependence --
  Even though the system size used here was not too small, for completeness an
  analysis of the effect of the system size should be done in
  order to have an idea of the smallest system
  sizes beyond which the variables of interest have similar statistical properties.
\item The effect of other kinds of boundary conditions can be explored -- there are
  many possibilities here.
\item Dynamics dependence -- Other than the Metropolis algorithm,
  many alternative dynamics protocols could be used to evolve the Ising
  configurations, such as Glauber dynamics\cite{glauber} and
  Kawasaki dynamics\cite{kawasaki}. If the results are qualitatively and/or
  quantitatively different, then a study on which of the dynamics is
  most appropriate should be done.
\item If possible, a comparison may be made to trends seen in some
  available data on knowledge spreading and look for the universal features
  in such data sets.
\item The need for promotion of the number of possible levels of knowledge could be explored.
  Two possible generalizations can be made.
  \begin{enumerate}
  \item Multiple areas or discrete levels of knowledge -- Potts model.\cite{potts}
  \item Multiple components of knowledge each having a continuous (real number) level
    -- $n$ component Heisenberg model, also known as the $O(n)$ model.\cite{on}
  \end{enumerate}
\item Finally, in a real situation, people are not arranged in a regular grid.
  The effect of removing the grid may also be looked at. This may be taken care of
  by studying a system with varying bond lengths or equivalently by adding disorder
  to the coupling constants.
\end{enumerate}

\subsection*{Acknowledgment}
The author would like to thank the Centre for Professional Development in Higher Education,
Delhi University and the other organizers of the Faculty Induction Programme that forced
the author to come up with an interdisciplinary research topic and obtain some preliminary results.
This may open a future research avenue for the author to explore.

\clearpage\pagebreak

\end{document}